\documentclass[aps,prl,reprint,nopacs,superscriptaddress]{revtex4-2}

\usepackage{epsfig,mathrsfs,xcolor,latexsym,subfigure,marginnote,graphicx,verbatim,relsize,mathrsfs,color,array,amsmath,amsfonts,amssymb,graphicx}
\usepackage{braket}
\usepackage[english]{babel}
\usepackage[utf8]{inputenc}
\usepackage[colorinlistoftodos, color=green!40, prependcaption]{todonotes}
\usepackage{amsthm}
\usepackage{mathtools}
\usepackage{physics}
\usepackage{xcolor}
\usepackage{graphicx}
\usepackage{adjustbox}
\usepackage{placeins}
\usepackage[T1]{fontenc}
\usepackage{lipsum}
\usepackage{csquotes}
\usepackage{float}
\usepackage{booktabs}
\usepackage{amsmath}
\usepackage{float}
\usepackage{array}
\usepackage[breaklinks=true,linkcolor=blue, pdftex, pdftitle={Article}, pdfauthor={Author}, colorlinks=true, citecolor=blue, urlcolor=blue, filecolor=blue]{hyperref}
\usepackage{cleveref}
\newcommand{\mr}[1]{\mathrm{#1}}

\newcommand{\mx}{\mathrm{x}}
\newcommand{\mc}{\mathrm{c}}
\newcommand{\mxc}{\mathrm{xc}}

\begin{document}
\title{%How Semilocal Are Semilocal Density Functional Approximations?\\ ---Tackling Self-Interaction Error in One-Electron Systems}
Tackling the One-Electron Self-Interaction Error within the Semilocal Density Functional framework}

\author{Akilan Ramasamy,$^1$ Lin Hou,$^{1,2}$ Jorge Vega Bazantes,$^1$ Tom J. P. Irons,$^3$ Andrew M. Wibowo-Teale,$^{3,4}$ Timo Lebeda,$^1$ and Jianwei Sun}
\email[Correspondence email address: ]{jsun@tulane.edu}
\affiliation{Department of Physics and Engineering Physics, Tulane University, New Orleans, Louisiana 70118, USA
\\$^2$Theoretical Division, Los Alamos National Laboratory, Los Alamos, New Mexico 87545, USA
\\$^3$School of Chemistry, University of Nottingham, University Park, Nottingham NG7 2RD, United Kingdom
\\$^4$Hylleraas Centre for Quantum Molecular Sciences, Department of Chemistry, University of Oslo, P.O. Box 1033,
N-0315 Oslo, Norway}
% \affiliation{$^2$Theoretical Division, Los Alamos National Laboratory, Los Alamos, New Mexico 87545, USA}

\date{\today} % Leave empty to omit a date

\begin{abstract}
One-electron self-interaction errors (SIE), caused by incomplete cancellation of the electron’s spurious self-Coulomb interaction, pose a persistent challenge in density functional approximations, as illustrated by the prototypical one-electron system $\mr{H}_2^+$. While significant efforts have been made to eliminate SIE through the development of computationally expensive nonlocal density functionals, it is equally important to explore whether SIE can be mitigated within the framework of more efficient semilocal density functionals. In this study, we present a nonempirical exchange-only meta-generalized gradient approximation (meta-GGA) that incorporates the Laplacian of the electron density. Our results demonstrate that this meta-GGA significantly reduces SIE, yielding a binding energy curve for $\mr{H}_2^+$ that matches the exact solution at equilibrium and improves across a broad range of bond lengths over those of the Perdew-Burke-Ernzerhof (PBE) and strongly-constrained and appropriately-normed (SCAN) semilocal density functionals. This advancement paves the way for further development within the realm of semilocal approximations.
%Modern density functionals often suffer from self-interaction errors, leading to inaccurate energy descriptions in one-electron systems. Meta-generalized gradient approximations (meta-GGAs) rely on ingredients such as electron density, density gradient, density Laplacian, or kinetic energy density. We introduce a meta-GGA functional based on laplacian designed to minimize self-interaction errors, achieving exact or nearly exact results for analytically known one-electron densities and significantly improving the binding energy curve of $H_2^+$. The inclusion of the density Laplacian, which serves as a more effective descriptor for bonded systems, especially at varying bond lengths, enables our functional to surpass other semilocal approximations in describing $H_2^+$.
\end{abstract}

% \keywords{}

\maketitle

Kohn-Sham density functional theory (KS-DFT) \cite{KS_paper} is a widely used quantum mechanical method in condensed-matter physics and quantum chemistry due to its balance of computational efficiency and accuracy. DFT reduces the complexity of many-electron systems by describing their ground-state properties in terms of the electron density rather than the many-body wavefunction \cite{HK_theorems}, thereby significantly reducing the computational cost. In principle, DFT is exact for any system of electrons under an external potential typically generated by nuclei, with the total electronic energy given by
\begin{equation}
    E[n] = T_\mathrm{s}[n] + E_\mathrm{H}[n] + E_\mathrm{en}[n] + E_\mxc[n],
\end{equation}
where $n(r)$ is the electron density, $T_\mr{s}[n]$ the kinetic energy, $E_\mr{H}[n]$ the Hartree energy of classical Coulomb repulsion between electrons, $E_\mr{en}[n]$ the electron-nuclei interaction, and $E_\mxc[n]$ the exchange-correlation (XC) energy. However, in practice, the XC  energy functional, $E_\mxc[n]$ = $E_\mx[n] + E_\mc[n]$, must be approximated, as its exact form in terms of $n(r)$ is not explicitly known.

Among the key challenges in widely used density functional approximations (DFAs) are self-interaction errors (SIE), that leads to the delocalization error~\cite{perdew1982density, PZSIC,johnson2023delocalization} in common materials modeling~\cite{hou2025artificial}. SIE can be classified into one-electron SIE~\cite{PZSIC} and many-electron SIE~\cite{sie_2,BDE,ruzsinszky2006spurious}. In any one-electron system, the exchange energy should exactly cancel the Hartree energy, \( E_\text{x}[n] = -E_\mr{H}[n] \), and the correlation energy should vanish, \( E_\text{c}[n] = 0 \), as there is no electron-electron interaction. The Hartree-Fock (HF) method satisfies these conditions exactly, making it an exact approach for one-electron systems. In contrast, typical DFAs fail to achieve this cancellation, leading to one-electron SIE due to the incomplete removal of the spurious classical self-Coulomb interaction. The Perdew–Zunger self-interaction correction (PZ-SIC)~\cite{PZSIC}, which removes SIE on an orbital-by-orbital basis, eliminates one-electron SIE, leading to the extension of DFT to fractional electron numbers and revealing the derivative discontinuity of the energy~\cite{perdew1982density}. Many-electron SIE, by contrast, is defined as the deviation from piecewise linearity in the total energy between adjacent integer electron numbers~\cite{perdew1982density, BDE, ruzsinszky2006spurious}. Although eliminating one-electron SIE does not guarantee the absence of many-electron SIE in DFA constructions, the two are related, and correcting one often offers a promising route to addressing the other ~\cite{BDE,sie_2,kronik2020piecewise}.
%With this definition, HF is not many-electron SIE free in general since its energy profile is usually concave between adjacent integer numbers. Recent studies have further shown that eliminating one-electron SIE does not guarantee the absence of many-electron SIE in DFA constructions~\cite{BDE,sie_2,kronik2020piecewise}. Nonetheless, the two are related, and correcting one often offers a promising route to addressing the other.

%{\color{cyan}TL: Try of a more positive wording:\\Although eliminating one-electron SIE does not guarantee the absence of many-electron SIE~\cite{BDE,sie_2,kronik2020piecewise} and vice versa \cite{kronik2020piecewise}, the two are related and correcting one often offers a promising route to addressing the other.}

In terms of performance, SIE in DFAs can lead to inaccuracies in predictions of properties such as bond dissociation energies, transition states, magnetic properties, and band gaps \cite{cohen2008insights, BDE, TS1, TS2, MP, BG1, BG2, zhang2020symmetry,zhang2019subtlety,zhang2022critical,fitzhugh2023comparative,zhang2025advances}. 
Over the years, various methods such as PZ-SIC and variations thereof \cite{PZSIC, pederson2014communication, zope2019step}, localized orbital scaling correction (LOSC) \cite{losc, mahler2022localized}, local hybrid density functionals \cite{Perdew2008,kirkpatrick2021pushing,Haasler2020}, and other nonlocal density functionals based on the exact exchange energy density \cite{Becke2013, Janesko2010,Kong2015} have been developed to reduce or eliminate SIE. Although these methods can offer improvements over typical semilocal DFAs, they come at a significantly increased computational demand due to their nonlocal nature.

\begin{figure}[H]
    \centering
    \includegraphics[width=1\linewidth]{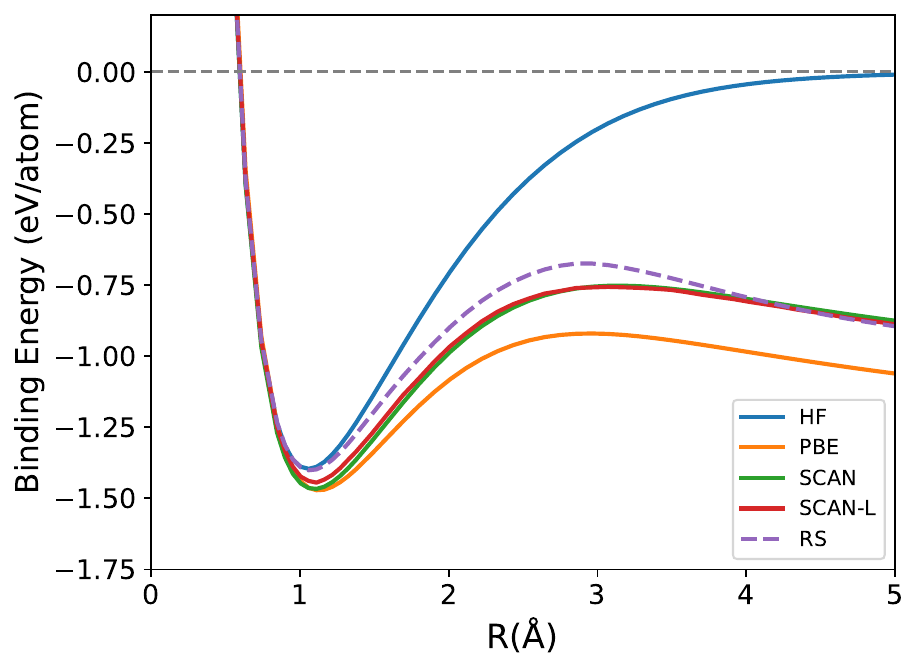}
    \caption{$\mr{H}_2^+$ binding energy curves of HF and exchange-only density functionals evaluated on HF orbitals.}
    %, using cc-pV5Z basis set.
    % \lh{LH: Also recommend to add more points around equilibrium. {\color{blue} JWS: I agree with Lin.}}}
    \label{fig:be curve}
\end{figure}

Therefore, it is equally important to evaluate how well SIE can be handled by computationally efficient semilocal DFAs. This is especially relevant given the success of the Strongly Constrained and Appropriately Normed (SCAN) \cite{SCAN} functional, which has shown significant improvements over the Perdew-Burke-Ernzerhof (PBE) \cite{pbe} functional across a wide range of materials and properties \cite{sun2016accurate}. SCAN achieves these advancements partially by reducing SIE within the semilocal framework \cite{zhang2020symmetry,zhang2019subtlety,zhang2022critical,fitzhugh2023comparative}, as shown in Fig.~\ref{fig:be curve} of the prototypical $\mr{H}_2^+$ binding energy curve, which is commonly used to define and illustrate one-electron SIE. This raises an intriguing question: can SIE be further minimized while retaining the computational efficiency of semilocal DFAs? Addressing this challenge could lead to the development of even more accurate and efficient functionals, expanding their utility across various applications.

In this letter, we present a positive answer to this critical question regarding one-electron SIE by introducing a non-empirical semilocal density functional, referred to as RS, depends on both gradient and Laplacian of the electron density.
As shown in Fig.~\ref{fig:be curve}, RS shows a clear improvement over popular semilocal DFAs such as PBE and SCAN in the $\mr{H}_2^+$ binding energy curve, demonstrating significantly reduced one-electron SIE in RS.

Semilocal DFAs can be written as,
\begin{equation}
    E_\mxc[n] = \int d^3\textbf{r}n\varepsilon_\mxc(n,\nabla n, \nabla^2 n, \tau).
\end{equation}
In the local density approximation (LDA) \cite{LSDA,lsda_1}, the electron density, $n(\textbf{r})$ is the only ingredient. 
Generalized gradient approximations (GGA) \cite{pbe,b88,gga_1} add the electron density gradient,
$\nabla n$. Meta-GGAs \cite{becke1989exchange, PKZB,TPSS,sun2012communication,SCAN,furness2020accurate, SCAN_L, mejia2020meta,kaplan2022laplacian, lebeda2024balancing} further include the Laplacian of electron density, $\nabla^2 n$, and/or the kinetic energy density, $\tau= \frac{1}{2}\sum_{i=1}^{N_\mathrm{occ}}|\nabla \phi_i|^2$, where $\phi_i$ are the Kohn-Sham orbitals. Meta-GGAs that depend on $\nabla^2 n$ are less explored, partially due to the complications in their potentials~\cite{mejia2020meta,kaplan2022laplacian,cancio2012laplacian} (see Section II in the Supplementary Material \cite{supplemental}, see also the references \cite{engel1992asymptotic,lindmaa2016energetics,kronik2020piecewise} therein). With more ingredients, semilocal DFAs can satisfy more exact constraints, i.e., known properties of the exact exchange-correlation energy \cite{SCAN,kaplan2023predictive,lebeda2025meta}.

SCAN is a meta-GGA that does not depend on $\nabla^2 n$ and uses \(\tau\) to identify different chemical environments through the variable \(\alpha = (\tau - \tau^\mathrm{W})/\tau^ \mathrm{unif}\) with $\tau^\mathrm{W} = |\nabla n|^2/8n$ and $\tau^ \mathrm{unif} = (3/10)(3\pi^2)^{2/3}n^{5/3}$ \cite{sun2013density}. 
The \(\tau\) dependence also allows SCAN to satisfy all 17 exact constraints suitable for semilocal functionals, including those for iso-orbital (one- and two-electron) systems that are not amenable to GGAs. Moreover, it provides the flexibility to guide the functional between exact constraints using appropriate norms, for example, the hydrogen atom, whose exchange-correlation holes are localized and deep \cite{SCAN}. %SCAN's improvement over the PBE GGA also stems from its ability to reduce self-interaction error (SIE), as illustrated in Fig.~\ref{fig:be curve}, further enhancing its accuracy.

%TL {\color{blue} Is this extensive discussion of SCAN-L required? I do not see the importance of comparing SCAN-L with SCAN for our approach to reducing one-electron SIE}
Meta-GGAs that depend on \(\nabla^2 n\) can be derived by de-orbitalizing \(\tau\)-dependent meta-GGAs, i.e., replacing \(\tau\) with a function of \(\nabla^2 n\), which offers faster computations. 
The SCAN-L functional \cite{SCAN_L} is a de-orbitalized version of SCAN, and it demonstrates surprisingly comparable accuracy for many systems and properties \cite{SCAN_L, mejia2018deorbitalized}, including the binding energy curve of \(\mr{H}_2^+\), as shown in Fig.~\ref{fig:be curve}. However, since \(\alpha\) is only approximated by the de-orbitalized \(\alpha_L(\nabla^2 n)\), SCAN-L only approximately satisfies many of the exact constraints that SCAN obeys. For instance, SCAN-L is not self-correlation free in one-electron systems. Furthermore, de-orbitalization does not guarantee that SCAN-L will be accurate for all appropriate norms. As shown in Tab.~\ref{exchange}, SCAN-L underestimates the magnitude of the exchange energy for the hydrogen atom.

For one-electron systems, \(\tau=\tau^\mathrm{W} = \frac{|\nabla n|^2}{8n}\), and the SCAN meta-GGA loses its dependence on the occupied orbitals $\phi_i$, and thus reduces to a GGA. In contrast, SCAN-L remains a meta-GGA, still dependent on \(\nabla^2 n\). Interestingly, despite this additional ingredient, SCAN-L does not improve over SCAN for the binding energy curve of \(\mr{H}_2^+\), and is even worse than SCAN for the exchange energies of two analytically known one-electron densities in Tab.~\ref{exchange}, the hydrogen atom (H) and the Gaussian (G) electron density. Thus, we can specify our question above as follows: can SIE in one-electron systems be further reduced by properly incorporating \(\nabla^2 n\) into the functional design, while still satisfying exact constraints and being guided by appropriate norms? Addressing this question would also provide important insight into reducing many-electron SIE \cite{sie_2,schmidt2016one} within the semilocal approximation framework, as the \(\tau\) dependence could be layered on top of the \(\nabla^2 n\) dependence at the meta-GGA level. 
%, which is fully determined by the electron density. Therefore, we omit the kinetic energy density as an ingredient in this work. Also, the correlation functional at the meta-GGA level can be made self-correlation free. We therefore only focus on the exchange density functional here.

To address this question, we start with the \(\nabla^2 n\)-dependent meta-GGA exchange energy of a spin-unpolarized density,
\begin{equation}
    E_\mx^\mathrm{MGGA}[n] = \int d^3\textbf{r}n\varepsilon_\mx^ \mathrm{unif}F_\mx(s,q)
\end{equation}
where $\varepsilon_\mx^\mathrm{unif} = -(3/4\pi)(3\pi^2n)$ is the exchange energy per particle of a uniform electron gas. $F_\mx(s,q)$ is the exchange enhancement factor depending on two dimensionless ingredients, the reduced density gradient $s = |\nabla n|/[2(3\pi^2)^{1/3}n^{4/3}]$
and the reduced density Laplacian $q = \nabla^2n/[4(3\pi^2)^{2/3}n^{5/3}]$.
The exchange energy for any spin-polarized density can be found from the exchange energy of a spin-unpolarized density using the exact spin-scaling relationship \cite{spin_scaling}. We do not consider correlation in this study, as $\tau$-dependent meta-GGAs are usually free from one-electron self-correlation \cite{TPSS,SCAN}.

For one-electron systems, SCAN reduces to the GGA 
\begin{equation}
    F_\mx^\mathrm{SCAN-i}(s)  = 1.174 \left(1 - e^{-as^{-1/2}}\right). 
    \label{eq:fxSCAN}
\end{equation}
Here, i in SCAN-i stands for iso-orbital systems. By design, $F_\mx^\mathrm{SCAN-i} \geq 0$, satisfying the exact constraint of a negative exchange energy \cite{primer}, $F_\mx^\mathrm{SCAN-i} \le 1.174$, satisfying the tight Lieb-Oxford bound for any one-electron density \cite{LO_bound}, and $F_\mx^\mathrm{SCAN-i} \propto s^{-1/2}$ as $s \rightarrow \infty$, satisfying the finite exchange energy per electron under the one-dimensional non-uniform coordinate scaling \cite{non_uni_cs}. 
The parameter $a$ = 4.9479 is then determined by the exchange energy of H. The resulting $F_\mx^\mathrm{SCAN-i}$ decreases monotonically with $s$, providing greater enhancement for electron densities with small $s$, which are typically more compact and slowly varying than those with large $s$.
%that are dilute and rapidly varying.
This simple construction works surprisingly well for one-electron systems, improving over PBE, as demonstrated in Tab.~\ref{exchange} for H and the Gaussian electron density and in Fig.~\ref{fig:be curve} for $\mr{H}_2^+$. It is thus desirable to maintain this simplicity when introducing a dependence on $q$ in $F_\mx(s,q)$. 

%\tl{TL: I moved the discussion of the Gaussian electron density before the definition of RS to enhance our construction strategy of adding a further appropriate norm for one-electron systems}

\begin{figure}[htb]
    \centering
    \includegraphics[width=1\linewidth]{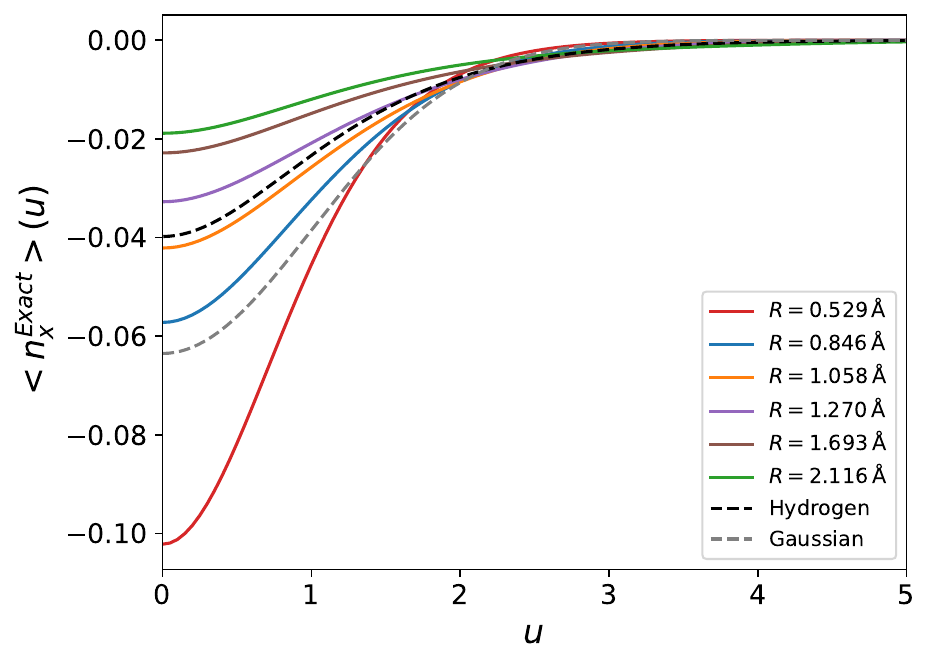}
    \caption{System- and spherically-averaged exact exchange holes versus the inter-electron distance $u$ for H, the Gaussian electron density, and $\mr{H}_2^+$ at various bond lengths $R$. 
    }
    \label{fig:xhole_plots}
\end{figure}

The additional dependence on $q$ allows us to satisfy more one-electron appropriate norms. Figure \ref{fig:xhole_plots} shows the system- and spherically-averaged exchange hole \cite{hou2024capturing,hou2024ab} versus the inter-electron distance $u$ for H and the Gaussian electron density along with $\mr{H}_2^+$ at various bond lengths.
Compared to H, the Gaussian electron density and $\mr{H}_2^+$ with bond lengths $R \leq 1.058$\AA\ have deeper exchange holes, while $\mr{H}_2^+$ with bond lengths $R \geq 1.270$\AA\ have shallower ones.
This suggests that the Gaussian electron density is as qualified for being an appropriate norm for semilocal density functionals as H is, since both are known analytically and not related to bonding. 
%\tl{TL: The following does, in my opinion, not relate to the paper:   The comparison between the PBE exchange holes and the exact ones given in section V of the supplementary material, shows that PBE's exchange holes are more short-ranged than the exact ones, and become too deep for large bond lengths R of $H_2^+$.}

It is interesting to note that although the exact exchange holes of the Gaussian electron density and of $\mr{H}_2^+$ at equilibrium ($R=1.058$ Å) are deeper than that of H, SCAN predicts a less negative exchange energy for the former (see Fig.~\ref{fig:be curve}) but an overly negative one for the latter (see Tab.~\ref{exchange}). This suggests that the SCAN-i GGA may be limited by its variables and could benefit from incorporating $q$ for improved accuracy.

\begin{center}
\begin{table}[H]
    \centering
    \small % Reduce font size by one point
    \setlength{\tabcolsep}{2pt} % Adjust column separation here
 \caption{Comparison of the exchange energies (in Ha) of different functionals for the H and Gaussian electron density.}
 
 \begin{tabular}{>{\small}c >{\small}c >{\small}c >{\small}c >{\small}c >{\small}c >{\small}c}
        \toprule
        \toprule
        System & LSDA & PBE & SCAN & SCAN-L & RS &  \textbf{Exact} \\
        \midrule
        Hydrogen & -0.2680 & -0.3059 & -0.3125 & -0.3110 & -0.3125 & -0.3125 \\
        Gaussian & -0.3410 & -0.3819 & -0.3975 & -0.3965 & -0.3989 & -0.3989 \\
        % Cuspless & -0.1050 & -0.1191 & -0.1224 & -0.1220 & -0.1228 & -0.1230 \\
        \bottomrule
        \bottomrule
    \end{tabular}
    
    \label{exchange}
\end{table}
\end{center}

In the following, we construct the RS enhancement factor for one-electron systems by satisfying exact constraints and using H and Gaussian electron density appropriate norms.
To keep the simplicity of SCAN's iso-orbital enhancement factor SCAN-i, we design our RS meta-GGA by coupling $F_\mx^\mathrm{SCAN-i}$ with a function $g(s,q)$,
\begin{equation}
    F_\mx^\mathrm{RS}(s, q) = 1.174 \left(1 - e^{-as^{-1/2}}\right)\cdot g(s,q),
\end{equation}
where
\begin{equation}
    g(s,q) = \left(\frac{1}{1 + \ln\left(1 + e^{b\cdot (q - q_0(s))}\right)}\right).%^{\frac{1}{m}}%
    \label{g2}
\end{equation}
The parameters $a$ and $b$ are determined by the appropriate norms of the H and Gaussian electron densities as discussed below. The function $q_0(s)$ that mixes the dependence of $q$ and $s$ is defined below. By construction, $0 \leq g(s,q) \leq 1$ for any $s \geq 0$ and $-\infty \leq q \leq \infty$. This guarantees that RS satisfies the exact constraints of a negative exchange energy and the tight Lieb-Oxford bound for one-electron densities. 

The function $g(s,q)$ is further designed to monotonically decrease with $q$, in view of the success of $F_\mx^\mathrm{SCAN-i}$ being monotonically decreasing with $s$ as discussed above. This requires $b > 0$. According to Bader's theory of atoms in molecules (AIM) \cite{AIM}, the behavior of \(\nabla^2 n\) provides critical insights into electronic structure. In regions of high electron density, such as at bond centers for compressed or equilibrium bonds and at nuclei, \(\nabla^2 n\) is negative. Conversely, in regions of low electron density, including inter-shell regions and bond centers of highly stretched bonds, \(\nabla^2 n\) becomes positive. For example, as the bond of $\rm{H}_2^+$ is stretched, $q$ at the bond center becomes more positive (see Section III in the Supplementary Material \cite{supplemental}).

% For an electron density under the one-dimensional nonuniform coordinate scaling, $q \to -\infty$ for some region as the coordinate scaling strength goes to $\infty$, and $q \to \infty$ for the other region (see Section IV in the Supplementary Material \cite{supplemental,non_uni_scaling,nush}).
% \ar{AR: 
% under one-dimensional nonuniform coordinate scaling in cylindrical coordinates, $q \to -\infty$ in regions where the system is highly compressed along the axial direction (i.e., small axial coordinate $u$, for any fixed radial distance $\rho$) while $q \to \infty$ throughout the remaining spatial domain as the coordinate scaling strength approaches $\infty$ (see Section IV in the Supplementary Material \cite{supplemental,non_uni_scaling,nush})}

For an electron density under one-dimensional nonuniform coordinate scaling along the axial direction represented in a cylindrical coordinate,
\( q \to -\infty \) in regions that have small axial distance from the origin at a fixed radial distance,
while \( q \to \infty \) elsewhere as the scaling strength approaches \(\infty\) (see Section IV in the Supplementary Material \cite{supplemental}, see also references \cite{non_uni_scaling,nush} therein).
In the region where $q \to -\infty$, $g(s,q) \to 1$, while for the other region, $q \to \infty$, $g(s,q) \to 0$. Therefore, $g(s,q)$ guarantees that $F_\mx^\mathrm{RS}$ reduces to $F_\mx^\mathrm{SCAN-i}$ in the $q \to -\infty$ region, and vanishes for the other region, yielding a finite exchange energy for one-electron systems under one-dimensional nonuniform coordinate scaling.

\begin{figure}[hbt]
    \centering
    \includegraphics[width=1\linewidth]{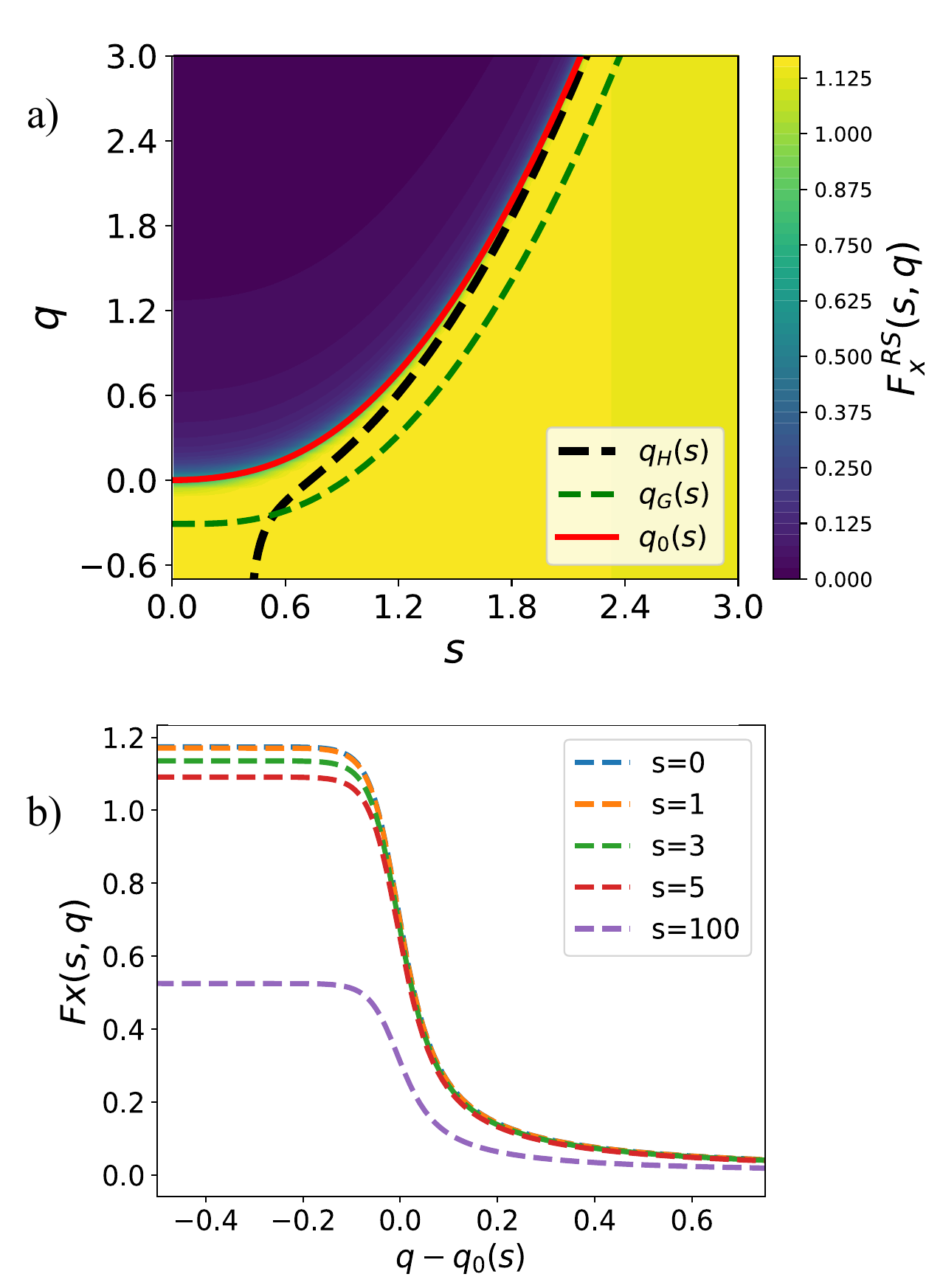}
    \caption{(a) Contour Plot of the RS enhancement factor $F_\mx^\mr{RS}$ versus s and q. 
    (b) $F_\mx^\mr{RS}$ as a function of ($q-q_0(s)$) for various values of s.
    }
    \label{contour_plots}
\end{figure}

To match the exact exchange energies of the appropriate norms, i.e., H and the Gaussian electron density, incorporating the $q$ dependence through $g(s,q)$ in the RS meta-GGA requires a careful design of the function $q_0(s)$. Due to the spherical symmetry and strict monotonicity of the electron densities, $q(r)$ of H and the Gaussian electron density can be expressed as functions of $s(r)$, denoted as $q_\mr{H}(s)$ and $q_\mr{G}(s)$ and shown in Fig.~\ref{contour_plots} (a) as black and green dashed lines, respectively. For most regions, $q_\mr{G}(s) < q_\mr{H}(s)$. 
$q_\mr{H}(s)$ has the analytical form \cite{supplemental}
\begin{equation} 
    q_\mr{H}(s) = s^2 \left[ 1 - \frac{2}{3 \ln{\left( (6\pi)^{1/3}s \right)}} \right]~.
\end{equation}
Our choice for $q_0(s)$ is a modification of $q_\mr{H}(s)$, 
\begin{equation} 
    q_0(s) = s^2 \left[ 1 - \frac{2}{3 \ln{\left( (6\pi)^{1/3}(1+s^2)^{1/2} \right)}} \right] \label{q0(s)} 
\end{equation}
by replacing $s$ in the denominator by $\sqrt{1+s^2}$. In doing so, $q_0(s)$ avoids the singularity of $q_\mr{H}(s)$ at $s=(6\pi)^{-1/3}$ and approaches $q_\mr{H}(s)$ asymptotically from above, as shown in Fig.~\ref{contour_plots} (a).
For $\mr{H}_2^+$, $q$ generally becomes more positive as the bond length $R$ increases (see Section III in the Supplementary Material \cite{supplemental}). Since $g(s,q)$ is inversely proportional to $q$, $F_\mx^\mathrm{RS}(s,q)$ therefore tends to decrease for larger $R$, as is desirable for a less negative exchange energy.

We determine the parameters \( a = 5.93 \) and \( b = 36.29 \) by fitting to the exact exchange energies of H and the Gaussian electron density (see Section I in the Supplementary Material \cite{supplemental}, see the references \cite{slsqp,scipy,psi4,Dunning1989,Woon1993,Woon1995,QUEST,Lebedev1992,Lebedev1976,ernzerhof1998generalized} for more details). This is reflected in RS matching the exact exchange energies of H and the Gaussian electron density in Tab.~\ref{exchange}.
Figure \ref{contour_plots}(b) illustrates that the enhancement factor decays rapidly with \( q \) for fixed \( s \) when crossing \( q_0(s) \). This is due to the large value of \( b \). As \( q_\mr{G}(s) - q_0(s) < -0.25 \) across most of the \( s \) range in Fig.~\ref{contour_plots}(a), the Gaussian electron density primarily experiences the form of \( F_\mx^\mathrm{SCAN-i}(s) \). Consequently, the parameter \( a \) is larger than in SCAN (where $a = 4.9479$), leading to a slower decay in \( s \) than in SCAN. This is a direct consequence of using the exact exchange energy of the Gaussian electron density as an additional appropriate norm.

\begin{figure}[htb]
    \centering
    \includegraphics[width=1\linewidth]{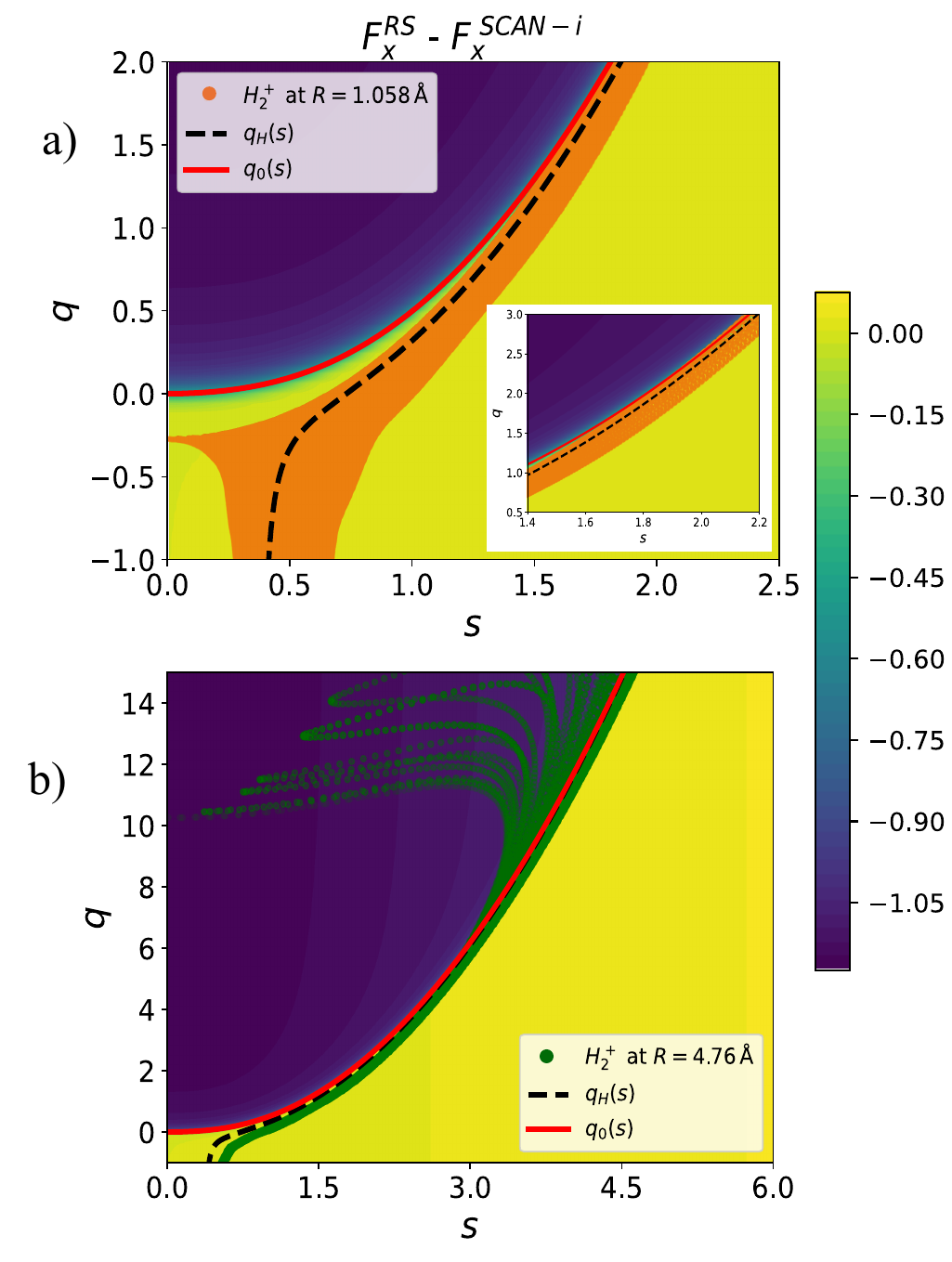}
    \caption{(a) Contour plot of the difference in the RS and SCAN-i enhancement factors against $s$ and $q$, overlaid by $(s, q)$ points existing in $\mr{H}_2^+$ at the equilibrium bond length ($R=1.058$\AA). The inset zooms in the range of $1.4 < s < 2.2$.  (b) Same as (a) but for $\mr{H}_2^+$ with $R=4.76$\AA.}
    \label{fig:diff_contour}
\end{figure}

Remarkably, the parameters obtained by fitting to H and the Gaussian electron density predict the total energy of \( \mr{H}_2^+ \) at equilibrium ($R = 1.058 \, \text{\AA}$) to be $-0.6026$ hartree, perfectly matching the exact HF value. Moreover, the RS binding energy curve aligns perfectly with the HF curve up to slightly beyond \( R = 1.058 \, \text{\AA} \). This consistency likely reflects the fact that the exchange holes of the Gaussian electron density and \( \mr{H}_2^+ \) (for \( R \leq 1.058 \, \text{\AA} \)) are deeper than those of hydrogen.
For stretched bond lengths between \( R = 1 \, \text{\AA} \) and \( R = 3 \, \text{\AA} \), the RS binding energy curve shows significant improvement over SCAN and other semilocal density functionals. This improvement highlights the importance of incorporating the reduced density Laplacian \( q \) in mitigating one-electron SIE.

To understand how RS improves over SCAN for the binding energy curve of \( \mr{H}_2^+ \), 
Fig.~\ref{fig:diff_contour} shows pairs $(s,q)$ occurring in $\mr{H}_2^+$ at equilibrium ($R=1.058 \AA$), overlaid on the contour plot of $F_\mx^\mathrm{RS} - F_\mx^\mathrm{SCAN-i}$. There are considerably many pairs $(s,q)$ that lie between the lines $q_0(s)$ and $q_\mr{H}(s)$, and some points with $s>1.5$ are even above the line $q_0(s)$, as shown in the inset. Given the rapid decay of $F_\mx^\mathrm{RS}$ versus $q - q_0(s)$ shown in Fig.~\ref{contour_plots} (b) and the exact match to the exchange energy of H, at these points $F_\mx^\mr{RS} < F_\mx^\mr{SCAN-i}$. Thus, the resulting energy of $\mr{H}_2^+$ at equilibrium is less negative for RS than for SCAN, and the RS energy perfectly matches the exact HF energy, as shown in Figure~\ref{fig:be curve}. As the bond distance increases, more pairs $(s,q)$ appear above $q_0(s)$, thereby improving the energy over that of SCAN.

However, for $R \gtrsim 4$ \AA, the improvement begins to diminish and ultimately RS becomes marginally worse than SCAN for $R>4$ \AA\ as shown in Fig.~\ref{fig:be curve}. This is because, as shown in Fig.~\ref{fig:diff_contour}(b), for $R=4.76$ \AA\ the pairs $(s,q)$ that lie above $q_0(s)$ have very large $q$ ($q > 6$) associated with relatively small $s$ and very low electron densities (see Section III in the Supplementary Material \cite{supplemental}). The energetically relevant electron density concentrates around the atomic centers with (s,q) below $q_H(s)$. Thus, the reduced enhancement of RS compared to SCAN occurs only in energetically irrelevant regions, and the slightly slower decay with $s$ of RS leads to a slightly larger overbinding than in SCAN. This is also dictated by the fact that for stretched bonds, as shown in Fig.~\ref{fig:xhole_plots}, the exchange holes are relatively shallow and more delocalized, making it difficult to approximate them with semilocal density functionals.

%{\color{red} JWS: Refine this part by referring to the supplementary materials. {\color{blue} AR: Further discussion and detailed analysis of RS’s improvement over SCAN can be found in the Supplementary Material. Notably, for bond lengths R > $4$\AA, RS performs slightly worse than SCAN, and the reasoning behind this behavior is provided in the Supplementary Material.} 

In summary, we have developed the nonempirical exchange-only $\nabla^2 n$-dependent RS meta-GGA to effectively reduce self-interaction errors in one-electron systems by satisfying all exact constraints for one-electron systems and being guided by suitable one-electron appropriate norms. The binding energy curve of $\mr{H}_2^+$, evaluated using Hartree-Fock orbitals, predicted by RS matches the exact HF one beyond the equilibrium point, significantly improving over SCAN across a wide range of bond lengths. 
The inclusion of $\nabla^2 n$ is shown to be crucial for reducing one-electron SIE in $\mr{H}_2^+$ at equilibrium and stretched bonds.

Although the RS meta-GGA itself was developed specifically for one-electron systems, this proof-of-principle is expected to lead to improved general purpose meta-GGAs by replacing the iso-orbital limit in a framework like SCAN by RS or a similar $\nabla^2 n$-dependent meta-GGA.
Our work therefore paves the way for further reducing SIE within the semilocal framework at the meta-GGA level while maximally satisfying suitable exact constraints with guidance from appropriate norms.

{\it Acknowledgments:} A.R. and J.S. acknowledge the
support of the U.S. Office of Naval Research (ONR)
Grant No. N00014-22-1-2673. L.H., J.V.B., and T.L. were supported by the National Science Foundation
(NSF) under Grant No. DMR-2042618, who carried out the calculations related to the exchange holes. The work at Los Alamos National Laboratory was carried out under the auspices of the U.S. Department of Energy (DOE) National Nuclear Security Administration under Contract No. 89233218CNA000001. It was supported by the LANL LDRD Program, and in part by the Center for Integrated Nanotechnologies, a DOE BES user facility, in partnership with the LANL Institutional Computing Program for computational resources. A.M.T. is grateful for support from the European Research Council under H2020/ERC Consolidator Grant “topDFT” (Grant No. 772259).

\bibliography{bib}

\end{document}

\title{Supplementary Material: Tackling the One-Electron Self-Interaction Error within the Semilocal Density Functional framework}

%How Semilocal Are Semilocal Density Functional Approximations?\\ ---Tackling Self-Interaction Error in One-Electron Systems

\author{Akilan Ramasamy,$^1$ Lin Hou,$^{1,2}$ Jorge Vega Bazantes,$^1$ Tom J. P. Irons,$^3$ Andrew M. Wibowo-Teale,$^{3,4}$ Timo Lebeda,$^1$ and Jianwei Sun}
\email[Correspondence email address: ]{jsun@tulane.edu}
\affiliation{Department of Physics and Engineering Physics, Tulane University, New Orleans, Louisiana 70118, USA
\\$^2$Theoretical Division, Los Alamos National Laboratory, Los Alamos, New Mexico 87545, USA
\\$^3$School of Chemistry, University of Nottingham, University Park, Nottingham NG7 2RD, United Kingdom
\\$^4$Hylleraas Centre for Quantum Molecular Sciences, Department of Chemistry, University of Oslo, P.O. Box 1033,
N-0315 Oslo, Norway}
% \affiliation{$^2$Theoretical Division, Los Alamos National Laboratory, Los Alamos, New Mexico 87545, USA}

\maketitle

\date{\today} % Leave empty to omit a date

% \keywords{}
\setcounter{figure}{0}
\renewcommand{\figurename}{Figure}
\renewcommand{\thefigure}{S\arabic{figure}}

\setcounter{equation}{0}
\renewcommand{\theequation}{S\arabic{equation}}

% \maketitle

%\section*{Supplementary Information}
%\label{sec:suppinfo}

\section{Computational Details}
For parameter optimization, we used the Sequential Least Squares Programming (SLSQP) \cite{slsqp} algorithm from the SciPy \cite{scipy} library. This algorithm offers the advantage of handling non-linear constraints, which is essential for satisfying the appropriate norms in our functional design. Specifically, we imposed a non-linear constraint to exactly recover the exchange energy of hydrogen, -0.3125 Ha. Additionally, the form of the functional was optimized to minimize the loss function for the Gaussian electron density. The loss function used for the Gaussian electron density is defined as,
\begin{equation}
\label{eq:lossgau}
    \mathcal{L}^\textrm{G} = (E_\textrm{x}^\textrm{RS} - E_\textrm{x}^\textrm{Exact})^2,
\end{equation}
where, $\mathcal{L}^\textrm{G}$ represents the loss for the Gaussian density, $E_x^{RS}$ is the exchange energy using the RS functional, and $E_x^{Exact}$ is the exact exchange energy of gaussian, which is -0.3989 Ha. This approach ensures that our functional is accurately constrained for both hydrogen and Gaussian densities, providing a robust optimization framework. 

All $H_2^+$ calculations were performed using Psi4 \cite{psi4}. The Hartree-Fock (HF) electron density and total energy were computed using the cc-pV5Z basis set from the Dunning family~\cite{Dunning1989,Woon1993,Woon1995}, with the basis functions in their uncontracted spherical Gaussian form. This HF density was then used in a non-self-consistent manner to evaluate the exchange energies for various density functionals discussed in the main text. The binding energy curves for $H_2^+$ were generated based solely on the exchange energies, excluding the correlation energy contributions for all the functionals considered.

The computational procedures for both the Hartree-Fock exact exchange hole and the PBE exchange hole model largely follow the methodologies described in Refs.~\cite{hou2024capturing,hou2024ab}, with calculations performed using the \textsc{Quest} code~\cite{QUEST}. The PBE exchange hole model employed the Hartree-Fock electronic density and its gradient, and a spin-unrestricted formalism was adopted throughout to extend the analysis to single-electron states. 

Spherically-averaged exchange holes were constructed via angular integration, using an order-41 Lebedev quadrature grid at each reference point~\cite{Lebedev1976,Lebedev1992}. The PBE exchange hole model~\cite{ernzerhof1998generalized} were computed with a spatial sampling interval of 0.01 bohr for the interelectronic distance $u$.

\section{Discussions on the exchange potential}
For self-consistent calculations, the exchange potential has to be implemented, 
\begin{equation}
    v_x = \frac{\partial e_x}{\partial n} - \nabla.\Bigg[\frac{\partial e_x}{\partial(\nabla n)} \Bigg] + \nabla^2.\Bigg[\frac{\partial e_x}{\partial(\nabla^2 n)} \Bigg]
\end{equation}
As the main focus in this article is to explore the potential of including $\nabla^2n$ in tackling the one-electron SIE at the energy level, we didn't implement $v_x$ for RS. 
The second term of the exchange potential can diverge at nuclei, while the third term involves the fourth-order derivatives of the electron density, potentially leading to numerical instabilities. However, since RS is designed such that it becomes independent of $q$ for $q<0$ and almost independent of $s$ for $s<3$, and since $q\to-\infty$ and $s=0.376$ at nuclei, our construction is not prone to this divergence. 

The third term, often denoted as the curvature term, can also potentially cause numerical instability~\cite{mejia2020meta,kaplan2022laplacian,cancio2012laplacian}. There are two sources for the numerical instability. One is the specific form of $e_x$ with respect to $q$, and the other is the involvement of fourth-order derivatives of the electron density. For example, the de-orbitalized q-dependent r$^2$SCAN-L metaGGA is more numerically stable than the de-orbitalized q-dependent SCAN-L metaGGA following their parent r$^2$SCAN and SCAN metaGGAs' trend and highlighting the importance of the functional form~\cite{mejia2020meta}.  
r$^2$SCAN-L has been shown to almost match the efficiency of the PBE GGA for the self-consistent field (SCF) cycles, but with a higher demand on the integration grid for solid simulations due to the involvement of fourth-order derivatives of the electron density ~\cite{mejia2020meta}. There has been a proposal to minimize the numerical instability caused by the curvature term in the potential by minimizing the curvature integral
\begin{equation}
     I = \int d^3r \left|\nabla \left(\frac{\partial e_\mathrm{xc}}{\partial \nabla^2 n} \right) \right|^2
\end{equation}
for some suitable test systems~\cite{cancio2012laplacian}.

The exchange potential should asymptotically decay as -1/r to exactly cancel the Hartree potential in the asymptotic limit for a finite system. However, it is known that semilocal approximations can not satisfy the asymptotic conditions on both exchange energy density $e_x(r)$ and $v_x(r)$ simultaneously~\cite{engel1992asymptotic}. Additionally, reproducing the exact asymptotic behavior of $v_x(r)$ was found to distort the quality of semilocal approximations in energetically more important regimes~\cite{engel1992asymptotic,lindmaa2016energetics}. In contrast, in many situations where one-electron SIE is important, e.g., the localized d- and f-electrons in transition metals, the asymptotic potential does not play any role. In this context, it is important to note that an asymptotically correct potential is not a necessary condition for a functional to be free from one-electron self-interaction~\cite{kronik2020piecewise}.

\section{Density ingredients plotted for different systems}
\begin{figure}[H]
    \centering
    \includegraphics[width=1\linewidth]{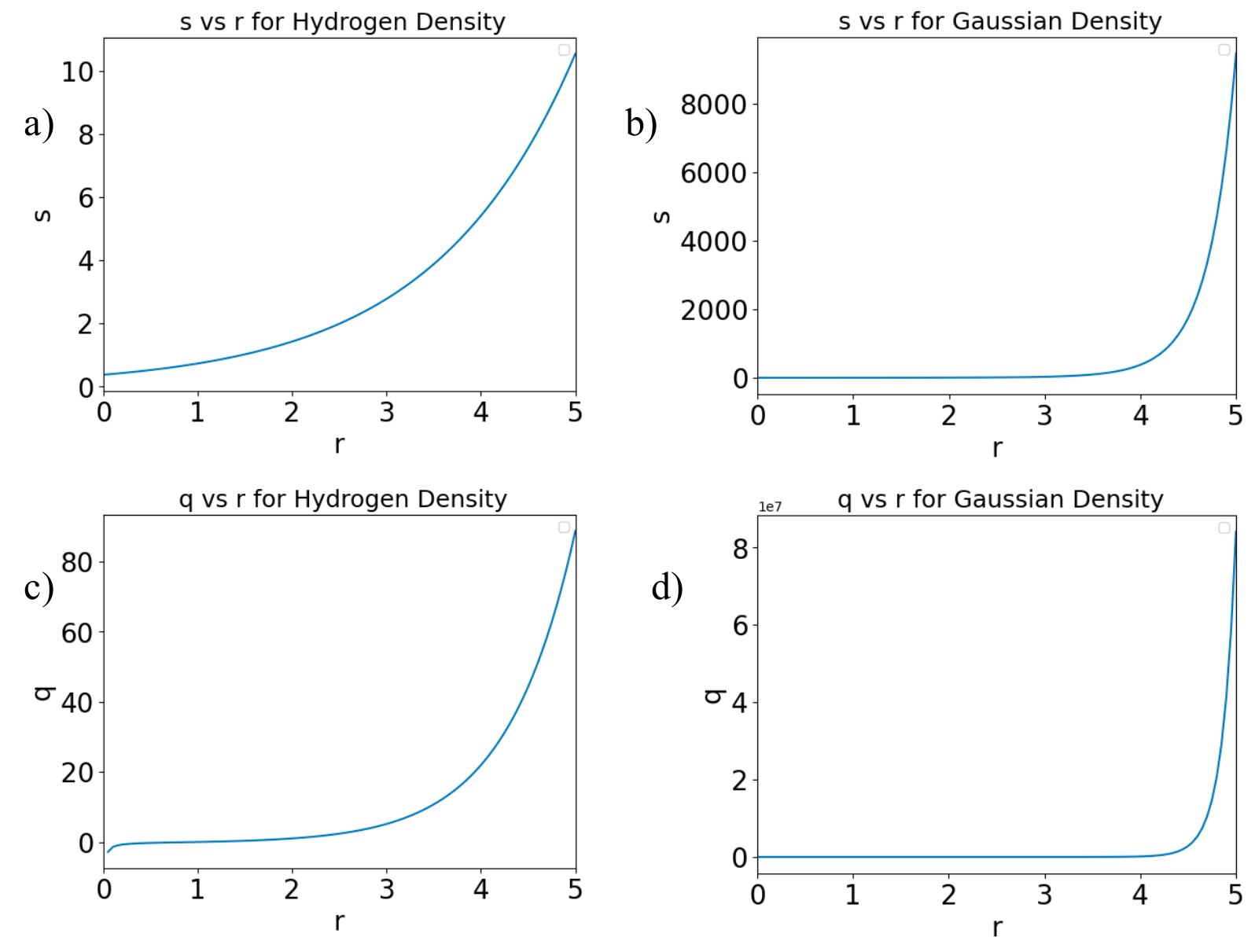}
    \caption{a) s vs r for hydrogen , b) s vs r for Gaussian electron density, c) q vs r for hydrogen, d) q vs r for Gaussian electron density (Note: Different scales are used for hydrogen and gaussian electron densities)}
    \label{fig:dens_ing}
\end{figure}

The density ingredients used in RS, s and q are plotted against position, r for the appropriate norms used, hydrogen density and gaussian density in Figure \ref{fig:dens_ing}.

\begin{figure}[H]
    \centering
    \includegraphics[width=1\linewidth]{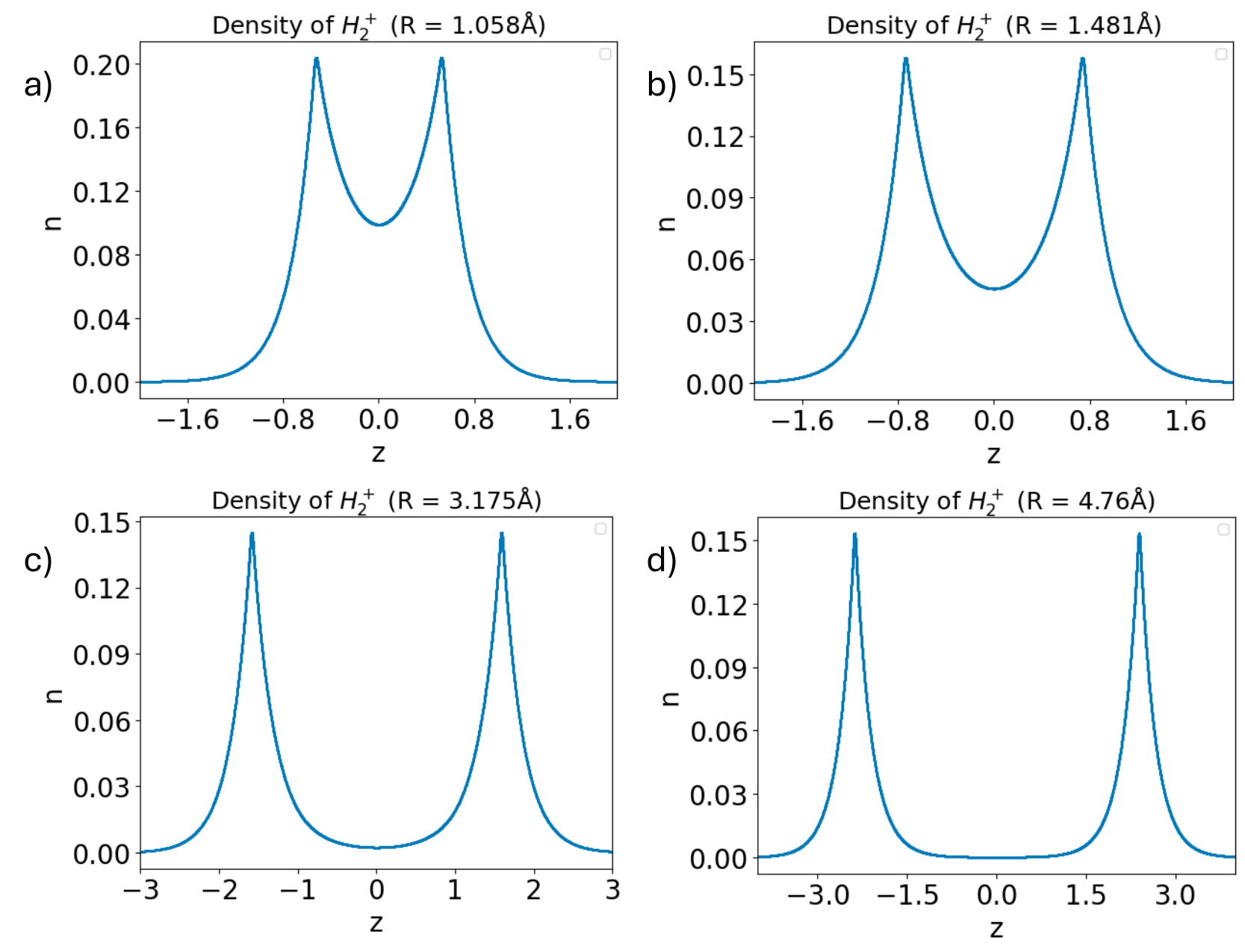}
    \caption{a) Electron density plotted against z for $R = 1.058$\AA , b) Electron density plotted against z for $R = 1.481$\AA, c) Electron density plotted against z for $R = 3.175$\AA, d) Electron density plotted against z for $R = 4.76$\AA}
    \label{fig:h2+_dens}
\end{figure}

As the bond length of $H_2^+$ increases, the electron becomes progressively more delocalized, leading to a significant reduction in electron density around the bond center. This trend is clearly illustrated in Figure \ref{fig:h2+_dens}, where the electron density diminishes as the nuclei move apart. 

\begin{figure}[H]
    \centering
    \includegraphics[width=1\linewidth]{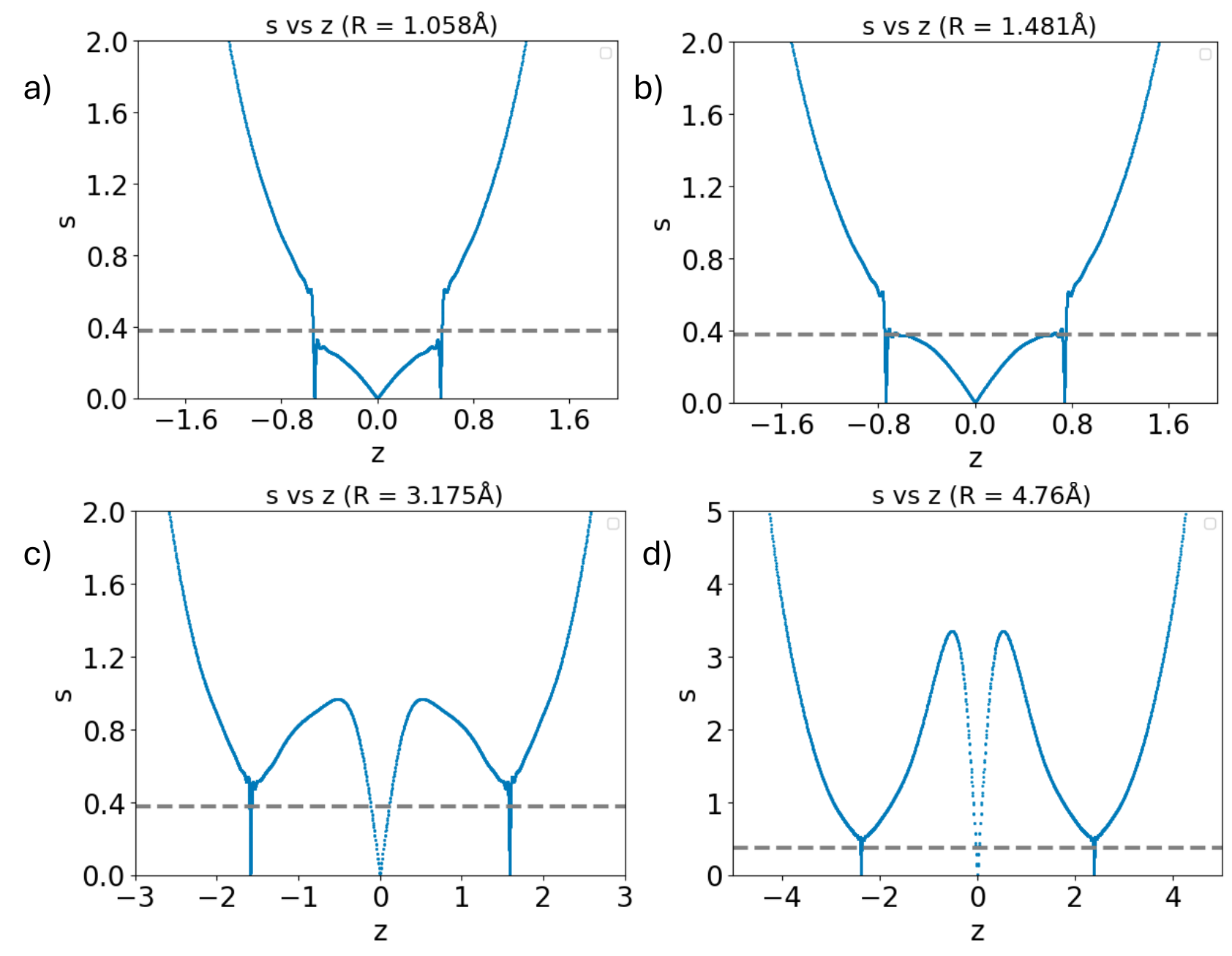}
    \caption{a) Reduced electron density gradient, s plotted against z for $R = 1.058$\AA , b) Reduced electron density gradient, s plotted against z for $R = 1.481$\AA, c) Reduced electron density gradient, s plotted against z for $R = 3.175$\AA, d) Reduced electron density gradient, s plotted against z for $R = 4.76$\AA}
    \label{fig:h2+_grad}
\end{figure}

At critical points, such as bond centers, the reduced electron density gradient, s consistently remains zero, regardless of the bond length, R. However, the evolution of 
s along the bond axis,
z varies as the bond length changes. Figure \ref{fig:h2+_grad} illustrates how 
s behaves in the direction of the bond for different bond lengths.

\begin{figure}[H]
    \centering
    \includegraphics[width=1\linewidth]{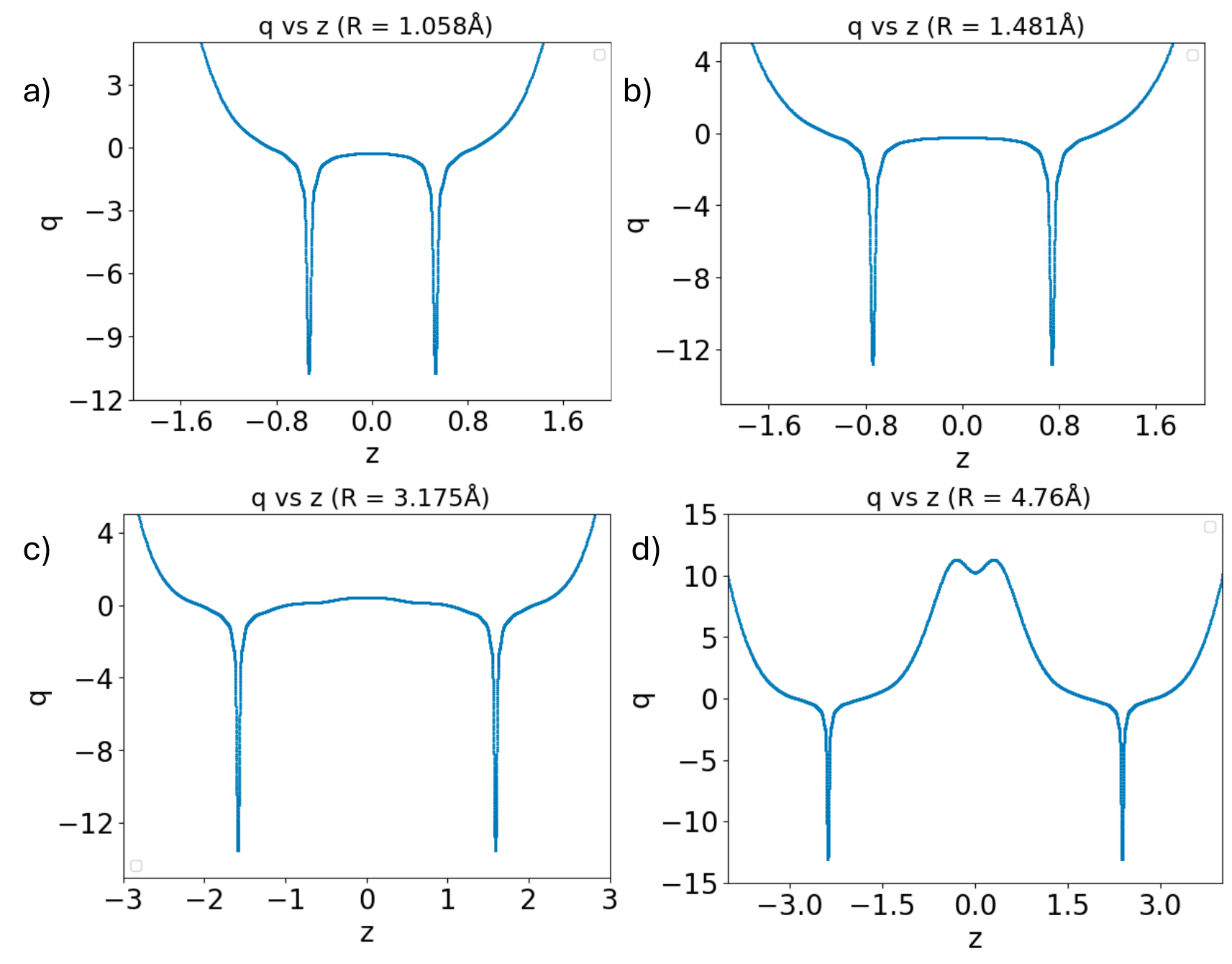}
    \caption{a) Reduced electron density Laplacian, q plotted against z for $R = 1.058$\AA , b) Reduced electron density Laplacian, q plotted against z for $R = 1.481$\AA, c) Reduced electron density Laplacian, q plotted against z for $R = 3.175$\AA, d) Reduced electron density Laplacian, q plotted against z for $R = 4.76$\AA}
    \label{fig:h2+_lapla}
\end{figure}

According to Bader's analysis \cite{AIM} as discussed in the main text, the reduced electron density Laplacian, q becomes increasingly positive as the electron density decreases, highlighting regions where the electron distribution is more diffuse. At the nuclear centers,
q diverges to $-\infty$, indicating the highly localized nature of the electron density in these regions. Figure \ref{fig:h2+_lapla} illustrates the behavior of
q along the bond axis,
z for different bond lengths. As the bond length increases, the electron density spreads out, leading to a corresponding rise in 
q at the bond center.

\section{Non-uniform coordinate scaling}
Electron density and Kohn-Sham orbitals that are non-uniformly scaled in one dimension (consider x direction in this case) is given by \cite{non_uni_scaling},
\begin{equation}
    n_\lambda^x(x,y,z) = \lambda n(u,y,z)
\end{equation}
\begin{equation}
    \phi_{i,\lambda}^x(x,y,z) = \lambda^{1/2} \phi_i(u,y,z)
\end{equation}

where $u = \lambda x$. The reduced density gradient, s is defined as,
\begin{equation}
    s = \frac{|\nabla n|}{2(3\pi^2)^{1/3}n^{4/3}}
    \label{s}
\end{equation}

Here, we define another dimensionless quantity $p = s^2$ for the simplicity of illustrating the scaling of s at the limits. When p is non-uniformly scaled along x direction, the dimensionless gradient p can be expressed as \cite{non_uni_scaling}

\begin{equation}
\begin{split}
    p_\lambda^x(x,y,z) = \frac{1}{4(3\pi^2)^{2/3}n(u,y,z)^{8/3}} \left( \lambda^{4/3} f_{p1}(u,y,z) \right. \\
    \left. + \lambda^{-2/3} f_{p2}(u,y,z) \right)
\end{split}
\end{equation}

where,
\begin{equation}
    f_{p1}(u,y,z) = [\frac{\partial n(u,y,z)}{\partial u}]^2
\end{equation}

\begin{equation}
    f_{p2}(u,y,z) = |\nabla_\perp n(u,y,z)|^2   
\end{equation}

When $\lambda \rightarrow 0,\ p$ will scale as $\lambda^{-2/3}$ and hence $s$ will scale as $\lambda^{-1/3}$. Similarly, when $\lambda \rightarrow \infty,\ p$ will scale as $\lambda^{4/3}$ and $s$ will scale as $\lambda^{2/3}$.

Similarly, we can define the reduced density laplacian, q as
\begin{equation}
        q = \frac{\nabla^2n}{4(3\pi^2)^{2/3}n^{5/3}}
        \label{q}
\end{equation}

When q is non-uniformly scaled along x direction, it can be expressed as,
\begin{equation}
\begin{split}
        q_\lambda^x(x,y,z) = \frac{1}{4(3\pi^2)^{2/3}n(u,y,z)^{5/3}}\left( \lambda^{4/3} f_{q1}(u,y,z) \right. \\
    \left. + \lambda^{-2/3} f_{q2}(u,y,z) \right)
\end{split}
\end{equation}

where,
\begin{equation}
    f_{q1}(u,y,z) = \frac{\partial^2 n(u,y,z)}{\partial u^2}
\end{equation}

\begin{equation}
    f_{q2}(u,y,z) = \nabla^2_\perp n(u,y,z)   
\end{equation}

\begin{figure}[H]
    \centering
    \includegraphics[width=1\linewidth]{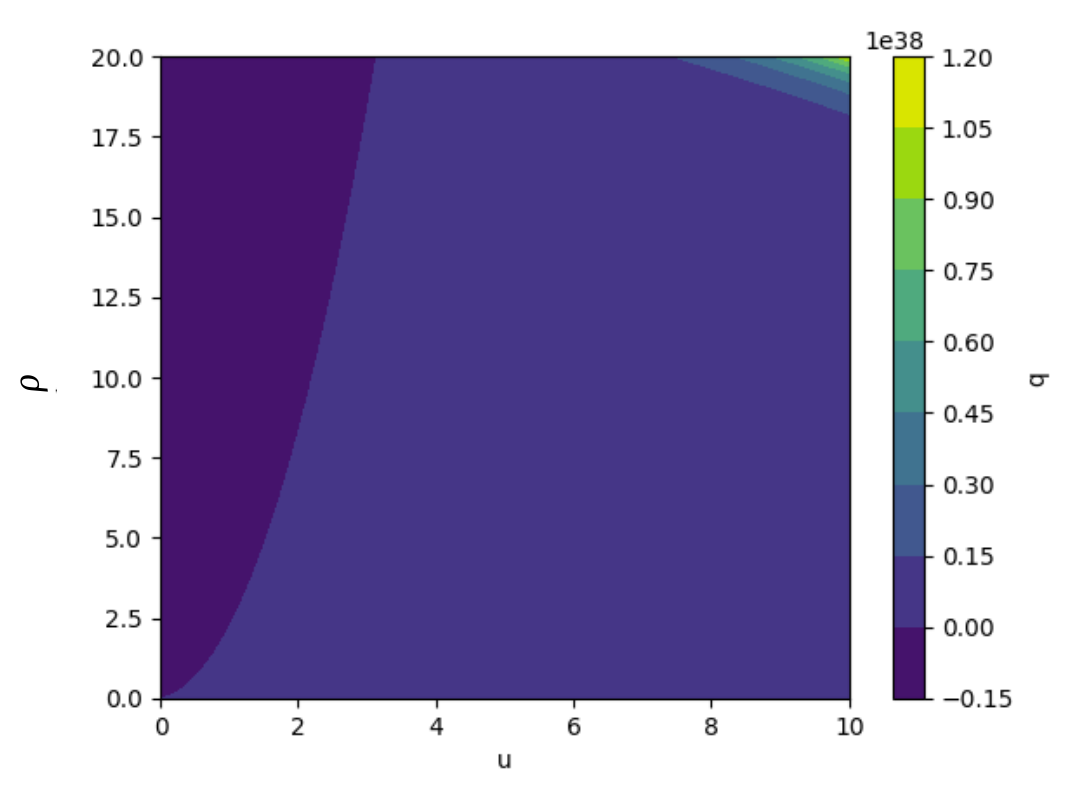}
    \caption{Contour plot of $q_\lambda^x$ against $u$ and $\rho$ for $\lambda=10^{20}$}
    \label{fig:q_nush}
\end{figure}

When $\lambda \rightarrow 0$, $q$ scales as $\lambda^{-2/3}$. In contrast, for $\lambda \rightarrow \infty$, $q$ scales as $\lambda^{4/3}$. In the latter case, $q$ may diverge to either $-\infty$ or $+\infty$, depending on the sign of $f_{q1}(u,y,z)$. Near or at the nucleus, where the second derivative of the density with respect to position $u$ ($f_{q1}(u,y,z)$) is negative, $q$ tends toward $-\infty$ as $\lambda \rightarrow \infty$. Conversely, in other regions where $f_{q1}(u,y,z)$ is positive, $q$ tends toward $+\infty$. Figure \ref{fig:q_nush} illustrates these limiting behaviors for $q$ as $\lambda \rightarrow \infty$, using the non-uniformly scaled hydrogen (NUSH) density \cite{nush}, with $q$ expressed in cylindrical coordinates ($u$ and $\rho$). 
As shown in the figure, for a large positive value of $\lambda$ (e.g., $10^{20}$), the value of $q$ scales down to approximately $-10^{38}$ in the small $u$ regions, while in other regions, it scales up to around $+10^{38}$.

The analytical expressions for the electron density, n, spin-scaled s and q derived for NUSH are given below,
\begin{equation}
    n_\lambda^x(r) = \frac{\lambda e^{-2\sqrt{(\lambda x)^2 + y^2 + z^2}}}{\pi}
\end{equation}

% \begin{equation} 
%     s[2 n_\lambda^x] = \frac{\left( \lambda^2 \left( \frac{\partial r_u}{\partial u} \right)^2 + \left( \frac{\partial r_u}{\partial y} \right)^2 + \left( \frac{\partial r_u}{\partial z} \right)^2 \right)^{1/2}}{(6\pi^2)^{1/3} (n_\lambda^x)^{1/3}}
% \end{equation}

\begin{equation}
    s\left[ 2 n_\lambda^x \right] = \frac{\sqrt{ \lambda^2 \left( \frac{\partial r_u}{\partial u} \right)^2 + \left( \frac{\partial r_u}{\partial y} \right)^2 + \left( \frac{\partial r_u}{\partial z} \right)^2 }}{(6\pi^2)^{1/3} \left( n_\lambda^x \right)^{1/3}}
\end{equation}

where $r_u = (u^2 + y^2 + z^2)^{1/2}$

\begin{equation}
    q[2 n_\lambda^x] = \frac{w(u,y,z,r_u,\lambda)}{4(6\pi^2)^{2/3}(n_\lambda^x)^{2/3}}
\end{equation}

where,

\begin{equation} 
\begin{split} 
    w(u, y, z, r_u, \lambda) = 2\lambda^2 \left( \frac{2u^2}{r_u^2} - \frac{1}{r_u} + \frac{u^2}{r_u^3} \right) \\ + 2 \left( \frac{2y^2}{r_u^2} - \frac{1}{r_u} + \frac{y^2}{r_u^3} \right) \\ + 2 \left( \frac{2z^2}{r_u^2} - \frac{1}{r_u} + \frac{z^2}{r_u^3} \right) 
\end{split} 
\end{equation}

% \section{More Plots on $F_x$}
% Explain these plots
% \begin{figure}[H]
%     \centering
%     \includegraphics[width=1\linewidth]{fx_plot_remain.png}
%     \caption{(a) Contour Plot of the enhancement factor of RS against s and q - Zoomed in (b) Enhancement factor of RS as a function of s for various values of q.}
%     \label{fig:fx_plot_remain}
% \end{figure}

% \section{Contour plot of $F_x^{SCAN} - F_x^{PBE}$}
% Remove the q0 and qh lines from it 

%%%% system and averaged exchange hole
\section{Derivation of the exact system- and
spherically-averaged exchange hole}
The analytic expression of the exact system- and
spherically-averaged exchange holes of the Hydrogenic and Gaussian electron densities can be derived.

For one-electron systems,
\begin{equation}
    n_x(\textbf{r},\textbf{r+u}) = -n(\textbf{r+u}).
\end{equation}

\subsection{Hydrogenic density}
Let us consider the normalized hydrogenic 1s orbital density:
\begin{equation}
   n^H(\mathbf{r})= \frac{\alpha^3}{8\pi} e^{-\alpha r}
   \label{hydrogen_dens}
\end{equation}

and find the spherically averaged exchange-hole function $n_{x}^{H}(\mathbf{r}, u)$ written as:
\begin{equation}
n_{x}^{H}(\mathbf{r}, u) = -\frac{1}{4\pi} \int d\Omega_\mathbf{u} n^{H}(\mathbf{r} + \mathbf{u}).
\end{equation}

Notice that $n_{x}^{H}(\mathbf{r}, u) = \frac{\alpha^3}{8\pi} e^{-\alpha |\mathbf{r}+ \mathbf{u}|} $ and $|\mathbf{r} + \mathbf{u}| = \sqrt{r^2 + u^2 + 2 r u cos(\theta)}$. Then, 
\begin{equation}
    n_x^H(\mathbf{r}, u) = - \frac{1}{4\pi} \int d\Omega_\mathbf{u} \frac{\alpha^3}{8\pi} e^{-\alpha \sqrt{r^2 + u^2 + 2 r u cos(\theta)}}
\end{equation}

where the solid angle is $d\Omega_\mathbf{u} = sin(\theta)d\theta d\phi$, with $\theta \in [0, \pi]$, and $\phi \in [0, 2\pi]$. Hence, after integrating we get:

\begin{equation}
\begin{split}
    n_x^H(\mathbf{r}, u) = -\frac{\alpha}{16\pi ru} \left((\alpha \abs{r - u}  + 1)e^{-\alpha\abs{r - u}} \right.\\ \left.
    - (\alpha \abs{r + u}  + 1)e^{-\alpha\abs{r + u}}
    \right)
\end{split}
\end{equation}
Next, we get the system- and
spherically-averaged exchange hole $\langle n_x^H \rangle(u)$:

% \begin{equation}
% \begin{split}
%     \langle n_x^H \rangle(u) = \frac{1}{N}\int d\mathbf{r} n^H(\textbf{r}) n_x^H(\mathbf{r}, u)
%     \\= 
%     -\frac{1}{N} \int dr r^2 sin(\theta)d\theta d\phi
%     \left(\frac{\alpha^3}{8\pi} e^{-\alpha r}\right) \\ %split here\\
%     \left(
%     \frac{\alpha}{16\pi ru} \left( (\alpha \abs{r - u}  + 1)e^{-\alpha\abs{r - u}} %split here
%     \right.\\ \left. 
%     -(\alpha \abs{r + u}  + 1)e^{-\alpha\abs{r + u}}\right)
%     \right)\\
%     = -\frac{\alpha ^4}{32 \pi  N u}
%     \Bigl(
%     \frac{1}{6} u^2 e^{-\alpha u} (\alpha  u+3)
%     + \\ .%split here 
%       \frac{e^{-\alpha u} (3 \alpha  u+2)}{4
%    \alpha ^2}
%    -\frac{e^{-\alpha u } (\alpha  u+2)}{4
%    \alpha ^2}
%     \Bigr)  
% \end{split} 
% \end{equation}

\begin{equation}
\begin{aligned}
    \langle n_x^H \rangle(u) &= \frac{1}{N} \int d\mathbf{r} \, n^H(\mathbf{r}) n_x^H(\mathbf{r}, u) \\
    &= -\frac{1}{N} \int dr \, r^2 \sin(\theta) \, d\theta \, d\phi \left( \frac{\alpha^3}{8\pi} e^{-\alpha r} \right) \\
    \times  \frac{\alpha}{16\pi r u}& \left[ (\alpha |r - u| + 1)  e^{-\alpha |r - u|}  - (\alpha |r + u| + 1) e^{-\alpha |r + u|}\right]  \\
    &= -\frac{\alpha^4}{32 \pi N u}
    \Biggl( \frac{1}{6} u^2 e^{-\alpha u} (\alpha u + 3) \\ 
    & \quad \quad + \frac{e^{-\alpha u} (3 \alpha u + 2)}{4 \alpha^2} - \frac{e^{-\alpha u} (\alpha u + 2)}{4 \alpha^2}
    \Biggr)
\end{aligned}
\end{equation}

% \begin{align}
%     \langle n_x^H \rangle(u) &= \frac{1}{N} \int d\mathbf{r} \, n^H(\mathbf{r}) n_x^H(\mathbf{r}, u) \notag \\
%     &= -\frac{1}{N} \int dr \, r^2 \sin(\theta) \, d\theta \, d\phi
%     \left( \frac{\alpha^3}{8\pi} e^{-\alpha r} \right) \notag \\
%     &\quad \times \left( 
%     \frac{\alpha}{16\pi r u} \left[ (\alpha |r - u| + 1) e^{-\alpha |r - u|} 
%     - (\alpha |r + u| + 1) e^{-\alpha |r + u|} \right]
%     \right) \notag \\
%     &= -\frac{\alpha^4}{32 \pi N u}
%     \Biggl(
%     \frac{1}{6} u^2 e^{-\alpha u} (\alpha u + 3)
%     + \frac{e^{-\alpha u} (3 \alpha u + 2)}{4 \alpha^2}
%     - \frac{e^{-\alpha u} (\alpha u + 2)}{4 \alpha^2}
%     \Biggr)
% \end{align}

Then, since the exchange energy is given by
\begin{equation}
    E_x = \int_{0}^{\infty} du 4\pi u^2 N \frac{\langle n_x^H \rangle(u)}{2u}
\end{equation}
we can check the correctness of our previous expressions by integrating over $u$ and setting $\alpha=2Z$ where $Z=1$ and $N=1$ to get $E_x = -0.3125 \; \textrm{Ha}$.

\subsection{Gaussian electron density}
In the same way as done for the Hydrogenic density, now let us consider the Gaussian electron density:
\begin{equation}
   n^{G}(r)= 
   \frac{e^{-r^2}}{\pi^{3/2}}.
\end{equation}

We can find the Gaussian spherically averaged exchange-hole function by:
\begin{align}
 n_{x}^{G}(\mathbf{r},u) &= -\frac{1}{4\pi}\int d\Omega_u n^G(\mathbf{r} + \mathbf{u})\\
 &= 
 -\frac{1}{4\pi}
 \int d\theta sin(\theta) \; d\phi \;  
 \frac{e^{-|r+u|^2}}{\pi^{3/2}}\\
 &=
  - \frac{1}{2 \pi^{3/2} ru}
e^{-(r^2 + u^2)} sinh(2ru)
 \end{align}

Next, we can get the system- and
spherically-averaged exchange hole $\langle n_x^G \rangle(u)$:
\begin{align}
    \langle n_x^G \rangle(u) &= \frac{1}{N}\int d\mathbf{r} n^G(\textbf{r}) n_X^{G}(\mathbf{r}, u)\\
    &= -\left(\frac{1}{N}\right) 
    \frac{e^{-\frac{u^2}{2}}}{2 \sqrt{2} \pi
   ^{3/2}}
\end{align}

Finally the exchange energy is obtained by integrating over $u$ and setting $N=1$:
\begin{align}
     E_x &= \int_{0}^{\infty} du \; 4\pi u^2 N \frac{\langle n_x^G \rangle(u)}{2u} = -0.3989 \; \textrm{Ha}
\end{align}
which confirms the consistency of the previous equations and the end of these proofs.
%%%% system and averaged exchange hole

\section{The deviations of PBE exchange holes from the exact ones in the one-electron systems}

\begin{figure}[H]
    \centering
    \includegraphics[width=1\linewidth]{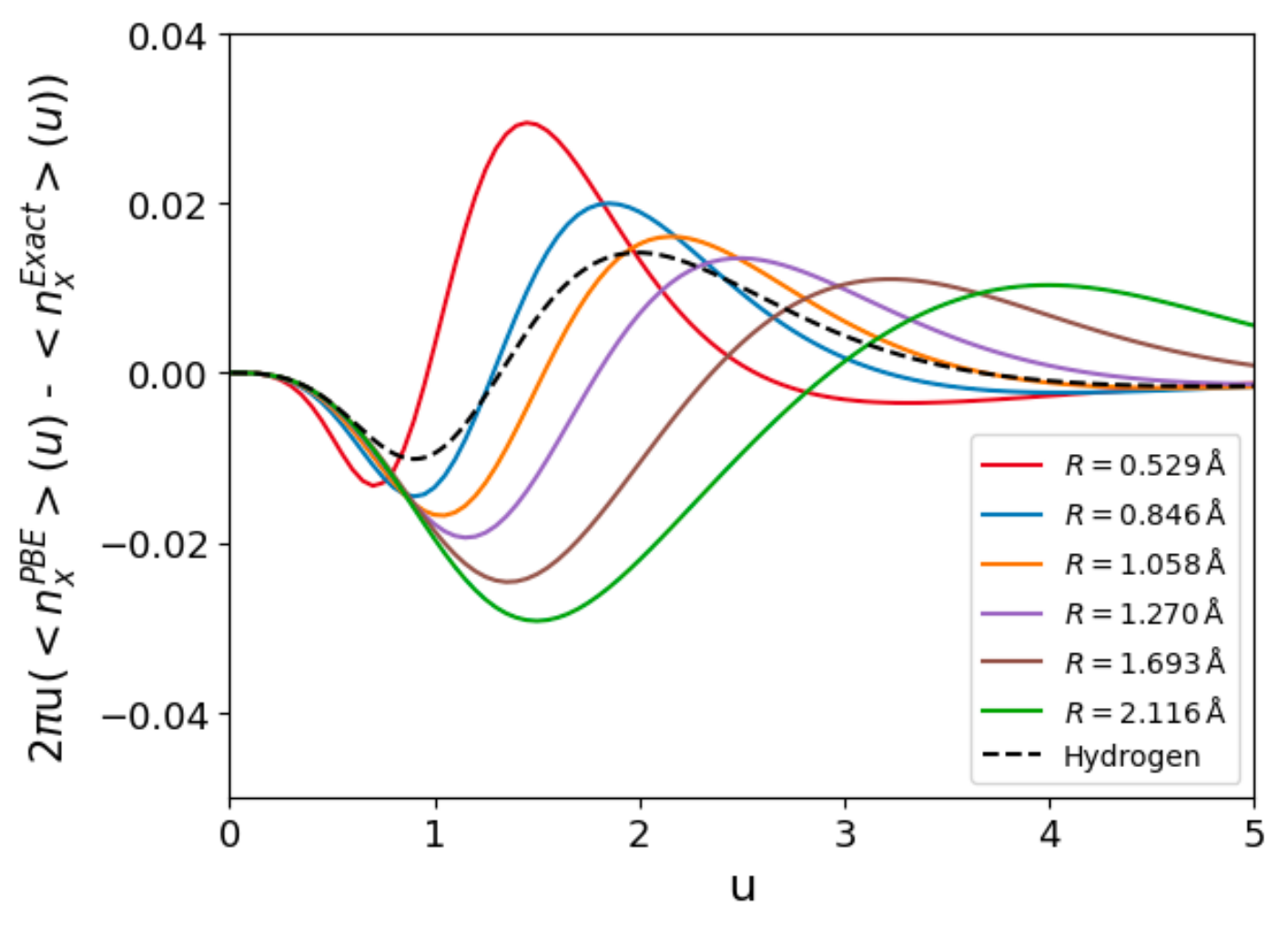}
    \caption{Difference between exact and the PBE system- and
    spherically-averaged exchange hole multiplied by 2$\pi$u plotted against u}
    \label{fig:xhole_diff}
\end{figure}

Figure \ref{fig:xhole_diff} plots the difference of PBE and exact system- and spherically-averaged exchange holes weighted by the inter-electron distance $u$. The integration of the above difference yields the deviation of PBE exchange energy from the exact one. It can be seen for all the considered systems that the PBE exchange holes are deeper in the short range than the exact ones, resulting in being shallower in the longer range due to the sum rule. For H and $H_2^+$ at small bond lengths ($R \leq 1.058 $\AA), the exact exchange hole is deep, and the overestimation in depth from the PBE exchange hole in the short range is small, followed by a localized peak representing an underestimation in hole depth. The cancellation between the well and the peak results in reasonably good accuracy for the exchange energies, for example, of H (-0.3059 Ha of PBE vs -0.3125 Ha of exact), although the PBE exchange hole was derived by enforcing the sum rule on the exchange hole of slowly varying densities \cite{ernzerhof1998generalized}. When the bond length R increases, the exact exchange hole becomes shallower as shown in Figure \ref{fig:xhole_diff} while the PBE exchange hole becomes too deep, leading to too negative exchange energies and large SIE, as shown in Figure 1 in the main text.

% However, GGAs are limited semilocal density functionals and the PBE GGA does not satisfy the exact constraints of iso-orbitals that are still suitable for semilocal approximations. SCAN, as a metaGGA, includes  the kinetic energy density as an ingredient, which allows SCAN to simultaneously satisfy all the 17 exact constraints and to design a specific GGA (eq. \ref{eq:fxSCAN}) as its component that satisfies the exact constraints of iso-orbital systems discussed earlier. Then, by making its exchange energy exact for the appropriate norm H, SCAN significantly improves over PBE for the Gaussian electron density exchange energy as shown in Table \ref{exchange} and the $H_2^+$ binding energy curve. 

Since SCAN satisfies exact constraints for one-electron systems that PBE does not and improves upon PBE in exchange energies, it is reasonable to believe that a properly reverse-engineered SCAN exchange hole model could provide shallower exchange holes than PBE, though still deeper than the exact ones. 
% Notably, the exact exchange holes of $H_2^+$ at equilibrium ($R=1.058$ Å) and the Gaussian electron density are deeper than those of H. However, SCAN predicts an overly negative exchange energy for the former (see Figure 1 in main text) and a less negative one for the latter (see Table 1 in main text). This suggests that the tailored GGA (eq. 4) may be limited by its variables and could benefit from incorporating $q$ for improved accuracy.

\section{Derivation of $q_H(s)$}
Let's substitute $\alpha = 2Z$ and $Z=1$ in Eq. \ref{hydrogen_dens}, the electron density for a hydrogen atom is then given by

\begin{equation} n^H(r) = \frac{e^{-2r}}{\pi}. \end{equation}

The spin-scaled reduced density gradient, $s$, can be derived from Eq.~\ref{s}, scaled by a factor of $\frac{1}{2^{1/3}}$ due to the spin scaling, and is expressed as

\begin{equation} s^H(r) = \frac{e^{-2r}}{(e^{-2r})^{4/3} (6\pi)^{1/3}}, \label{s_h} \end{equation} while the spin-scaled reduced electron density Laplacian, $q$, follows from Eq.~\ref{q} and is given by

\begin{equation} q^H(r) = \frac{e^{4r/3} \left(1-\frac{1}{r}\right)}{(6\pi)^{2/3}}. \label{q_r} \end{equation}

From Eq.~\ref{s_h}, $r$ can be expressed in terms of $s$ as

\begin{equation} r = \frac{3}{2}\ln\left( (6\pi)^{1/3}s \right). \label{r_s} \end{equation}

Substituting Eq.\ref{r_s} into Eq.\ref{q_r}, $q$ can then be expressed as a function of $s$, yielding the following relationship:

\begin{equation} q_H(s) = \left[1 - \frac{2}{3\ln\left((6\pi)^{1/3}s\right)}\right]s^2. \end{equation}

This formulation allows for $q_H(s)$ to be computed directly in terms of the reduced density gradient $s$ for the hydrogen atom.

\par
\section{Additional contour plot of $F_x^{RS} - F_x^{SCAN-i}$}

\begin{figure}[H]
    \centering
    \includegraphics[width=1\linewidth]{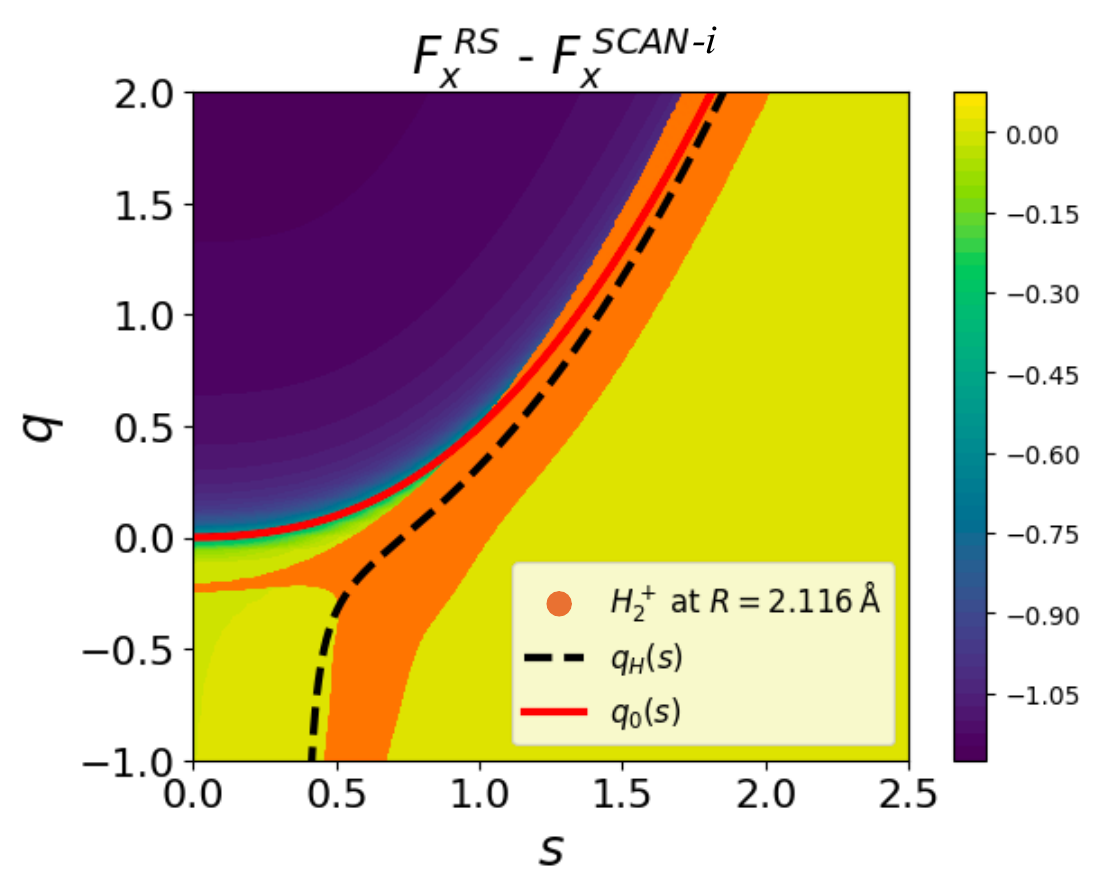}
    \caption{Contour Plot of the difference in the RS and SCAN enhancement factors against s and q, overlaid by (s, q) points existing in $H_2^+$ with bond length $R=2.116$\AA.}
    \label{fig:fx_diff_2.114}
\end{figure}

As bond lengths increases, more (s, q) points appear above $q_0(s)$, leading to improved energy descriptions compared to SCAN. To illustrate this, the (s,q) points at R = 2.116 \AA\ are plotted against the difference in enhancement factors between RS and SCAN in Figure \ref{fig:fx_diff_2.114}, in comparison with the plot of R = 1.058 \AA\ shown in Figure 4(a) of the main text. Notably, for regions where s > 1, a great number of data points lie above $q_0(s)$, resulting in a significantly smaller $F_x$ value relative to SCAN. This reduction in $F_x$ contributes to a more accurate energy description for $H_2^+$, improving the overall binding energy curve. 

\bibliography{bib}